# Metasurface Sensing Approach to DOA Estimation of Coherent Signals

Yishuo Zhao, Yan Hu and Yougen Xu

*Abstract*—The DOA estimation method of coherent signals based on periodical coding metasurface is proposed. After periodical coding, the DOA information of incident signals in the time domain is represented as the amplitude and phase information at different frequency points in the frequency domain. Finite time Fourier transform (FTFT) is performed on the received signal and appropriate frequency points are selected to reconstruct the frequency domain snapshot, then pattern smoothing (PS) technique is applied to execute DOA estimation. Compared with conventional DOA estimation methods, the proposed method has two main advantages: one is that only a single receiving channel is needed to avoid the appearance of channel mismatch errors, the other is that it can process with multiple coherent signals. The performance curves of the proposed method are analyzed under different conditions and compared with existing methods. Simulation results show the effectiveness of the proposed method.

## I. Introduction

Direction of arrival (DOA) estimation is one of the main research directions in array signal processing, which has a wide range of applications in radar, communication, navigation and other fields [1]-[4]. Scholars have conducted in-depth researches on it, and established a relatively mature algorithm system. Conventional algorithms can be divided into the following categories, one is the beam scanning methods, such as conventional beamforming (CBF) method [5], one is the subspace based methods, most famous for multiple signal classification (MUSIC) method [6] and estimating signal parameters via rotational invariance technique (ESPRIT) [7], the last one is the fitting methods, such as maximum likelihood (ML) method [8]. Subspace based methods are the most commonly used methods which can achieve ultra-high precision resolution at present, but require the array systems have multiple sensors and corresponding receiving channels to receive signals, therefore, high cost and complex feeding system are the disadvantages, and it cannot be applied to portable or cost-effective equipment as well as the rapidly evolving information technologies such as fifth-generation (5G) mobile communication [9] and Internet of Things (IoT) [10].

In the past few decades, metasurface which consisted of periodic or quasi-periodic subwavelength meta-atoms have received much attention and is a breakthrough technology [11], [12]. By integrating active devices such as diodes, graphene, microelectromechanical systems (MEMS) and liquid crystal [13]-[16], metasurface can control the amplitude, phase, polarization and other characteristics of incident electromagnetic waves in different manners [17] and has been applied in many scenarios, such as synthetic aperture radar, reconfigurable beamforming, wireless communication [18]-[20] and so on. This kind of metasurface is programmed by different coding sequences and called as coding metasurface. By combining coding metasurface and array signal processing, that is, the signals in free space are received by the metasurface, and the reflected or transmitted signals of the metasurface are received by a single antenna, then only a single receiving channel is needed, and the complexity of array system can be reduced, the amplitude-phase errors caused by channel mismatch can be avoided, which is more suitable for the development of information technology. Therefore, there are some DOA estimation methods using metasurface are studied in recent years [21]-[24]. In general, the study of metasurface based DOA estimation is in the ascendant and deserves further study.

In practice, due to factors such as multipath reflection or human interference, the signals received by the antenna array may be coherent with each other [25], [26]. At present, the coherent incident signals are not considered in all the studies of DOA estimation methods using metasurface, so it is significant to study the estimation method in this case. This paper presents a DOA estimation method of coherent signals based on periodical coding metasurface and pattern smoothing (PS) technique, which has the following advantages. First, the use of metasurface can realize single-channel reception, reduces the system cost, and avoids amplitude-phase errors caused by channel mismatch. Second, the use of PS technique makes this method enable to deal with coherent signals, and it is still applicable in the case of incoherent incidence, which is more suitable for the actual scenarios than the existing methods.

The remaining of the paper is organized as follows. The second part introduces the flow of receiving and processing signals briefly, and puts forward the periodical coding pattern and a new metasurface signal model, then the PS technique and the DOA estimation algorithm is introduced in the third part. The next part states the simulation results, the difference of

This work was supported in part by the National Natural Science Foundation of China under Grants 62371042 and the Municipal Natural Science Foundation of Beijing under Grants 4222015.

The authors are with the School of Integrated Circuits and Electronics, Beijing Institute of Technology, Beijing 100081, China (e-mail: yszhao@bit.edu.cn; yanhu@bit.edu.cn; yougenxu@bit.edu.cn).

probability of resolution (PR) and root mean square error (RMSE) between the proposed method and the existing methods is compared in the case of multiple incident signals and single incident signal. The last part summarizes the full text.

## II. A New Metasurface Signal Model

The flow chart of receiving and processing signals is illustrated in Fig. 1. The metasurface is controlled by the coding unit, and a receiving antenna is placed at the $z$ axis to receive signal. The received signal will be demodulated to obtain the baseband signal, then the frequency snapshot can be extracted by applying finite time Fourier transform (FTFT) and selecting frequency points. Finally, PS technique is used to perform DOA estimation. This section describes the periodical coding pattern and deduces a new frequency domain model of the received signal.

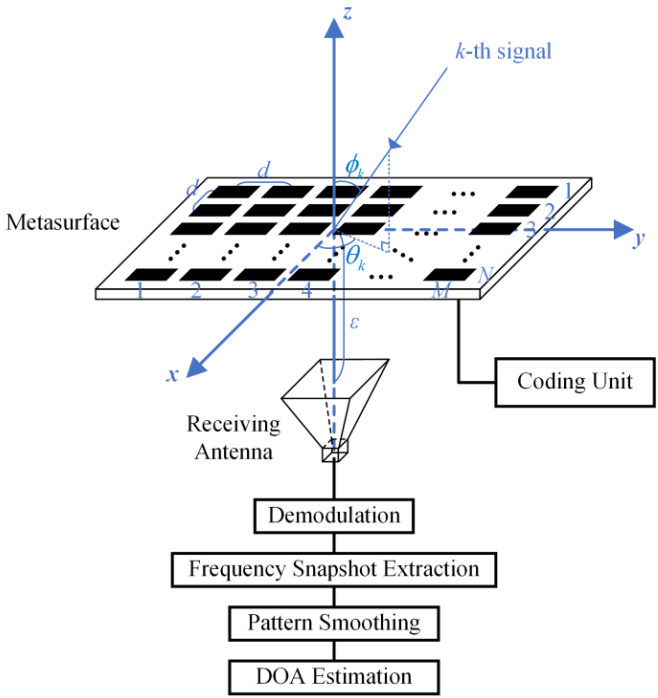

Fig. 1. The flow chart for receiving and processing signals, the blue line represents the rectangular coordinate system.

The metasurface employed consists of $M \times N$ uniformly spaced elements. The element spacing $d$ is the half of wavelength. The reference point is assumed to be the center of the metasurface, and the coordinate of the $(m,n)$-th element is

$$\boldsymbol{d}_{m,n} = \begin{bmatrix} n-(N+1)/2 \\ m-(M+1)/2 \\ 0 \end{bmatrix} d, \; n=1, 2, \cdots, N; \; m=1, 2, \cdots, M \quad (1)$$

The receiving antenna is located at $\boldsymbol{\varepsilon} = [0,0,-\varepsilon]^{\mathrm{T}}$, with $\varepsilon > 0$, as shown in Fig. 1.

The following periodical coding pattern is proposed:

$$w_{m,n}(t) = \begin{cases} 1 & \mu_1 \Delta T < t \leq \mu_2 \Delta T \\ -1 & i\Delta T < t \leq \mu_1 \Delta T, \; \mu_2 \Delta T < t \leq (i+1)\Delta T \end{cases} \quad (2)$$

where $\Delta T$ is the coding period, $i$ is an integer, and

$$\mu_1 = i + (m-1)/M + (n-1)/(MN) \quad (3)$$
$$\mu_2 = i + (m-1)/M + n/(MN) \quad (4)$$

the corresponding pattern sequence diagram is shown in Fig. 2.

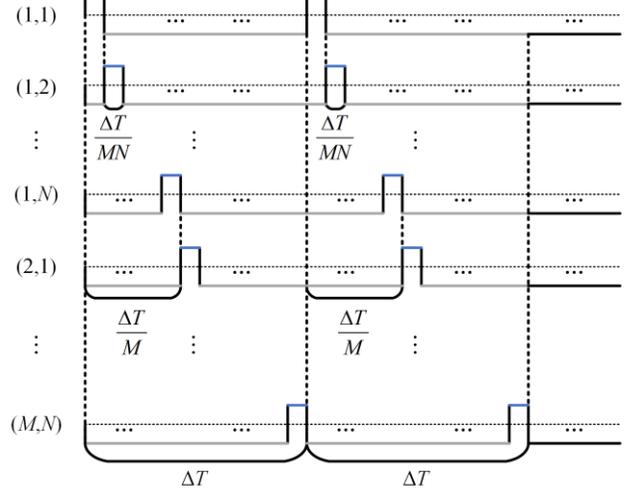

Fig. 2. The pattern sequence diagram corresponds to (2). The blue and grey lines represent 1 and −1 respectively. There are only two periods drawn.

With $K$ narrowband signals from far-field sources, incoherent or coherent, impinging upon the metasurface, the complex valued baseband output can be written as

$$x(t) = \sum_{m=1}^{M}\sum_{n=1}^{N}\sum_{k=1}^{K} w_{m,n}(t) e^{j\omega_0(\tau_{m,n,k}+\tau_{m,n})} s_k(t) + n(t) \quad (5)$$

where $\omega_0$ is the carrier frequency, $j=\sqrt{-1}$, $s_k(t)$ is the complex amplitude of the $k$-th signal, $n(t)$ is the additive noise, and

$$\tau_{m,n,k} = \boldsymbol{d}_{m,n}^{\mathrm{T}} \boldsymbol{k}_k / c \quad (6)$$
$$\tau_{m,n} = |\boldsymbol{d}_{m,n} - \boldsymbol{\varepsilon}|/c \quad (7)$$
$$\boldsymbol{k}_k = [\sin\phi_k \cos\theta_k, \sin\phi_k \sin\theta_k, \cos\phi_k]^{\mathrm{T}} \quad (8)$$

in which $c$ is the propagation speed of signal waves, "$|\cdot|$" stands for modulus, and $(\theta_k, \phi_k)$ is the azimuth-elevation 2-D DOA of the $k$-th signal.

We note that $w_{m,n}(t)$ is a periodical function of $t$ with period $\Delta T$; therefore, it has the following form of Fourier series:

$$w_{m,n}(t) = \sum_{p=-\infty}^{\infty} u_{m,n,p} e^{j2\pi \frac{p}{\Delta T} t} \quad (9)$$

where $u_{m,n,p}$ is the Fourier coefficient, given by

$$u_{m,n,p} = \begin{cases} 2(\mu_2 - \mu_1) - 1, & p=0 \\ \sum_{i=1}^{3} c_{1,i} \varsigma_i \mathrm{Sa}(p\pi c_{1,i}) e^{-jp\pi c_{2,i}}, & p \neq 0 \end{cases} \quad (10)$$

in which $\mathrm{Sa}(x) = \sin(x)/x$, $\varsigma_1 = \varsigma_3 = -1$, $\varsigma_2 = 1$, and

$$c_{1,1} = c_{2,1} = (m-1)/M + (n-1)/(MN) \quad (11)$$
$$c_{1,2} = 1/(MN) \quad (12)$$
$$c_{2,2} = 2(m-1)/M + (2n-1)/(MN) \quad (13)$$

$$c_{1,3} = (M-m+1)/M - n/(MN) \quad (14)$$

$$c_{2,3} = (M+m-1)/M + n/(MN) \quad (15)$$

Substitute (9) into (5), $x(t)$ can be rewritten as

$$x(t) = \sum_{k=1}^{K}\sum_{p=-\infty}^{\infty} \boldsymbol{u}_p^T \boldsymbol{a}_k s_k(t) e^{j2\pi \frac{p}{\Delta T} t} + n(t) \quad (16)$$

where $\boldsymbol{a}_k$ is the steering vector of the metasurface and $\boldsymbol{u}_p$ is the coefficient matrix, shown as

$$\boldsymbol{a}_k = [\boldsymbol{a}_{1,k}^T, \boldsymbol{a}_{2,k}^T, \cdots, \boldsymbol{a}_{M,k}^T]^T \quad (17)$$

$$\boldsymbol{a}_{m,k} = [e^{j\omega_0(\tau_{m,1,k}+\tau_{m,1})}, e^{j\omega_0(\tau_{m,2,k}+\tau_{m,2})}, \cdots, e^{j\omega_0(\tau_{m,N,k}+\tau_{m,N})}]^T \quad (18)$$

$$\boldsymbol{u}_p = [\boldsymbol{u}_{1,p}^T, \boldsymbol{u}_{2,p}^T, \cdots, \boldsymbol{u}_{M,p}^T]^T \quad (19)$$

$$\boldsymbol{u}_{m,p} = [u_{m,1,p}, u_{m,2,p}, \cdots, u_{m,N,p}]^T \quad (20)$$

Assume that the sampling time is $T_0$, the complex amplitude of the $k$-th signal remains unchanged during this time because of the characteristic of narrowband signals and its value is denoted as $s_k(t_0)$, consider the FTFT of the signal part of $x(t)$ in $T_0$, there is

$$\begin{aligned}\text{FTFT}(x(t)) &= \frac{1}{T_0}\int_0^{T_0}\sum_{k=1}^{K}\sum_{p=-\infty}^{\infty} \boldsymbol{u}_p^T \boldsymbol{a}_k s_k(t) e^{j2\pi\frac{p}{\Delta T}t} e^{-j\omega t} dt \\ &= \sum_{p=-\infty}^{\infty}\sum_{k=1}^{K} \frac{s_k(t_0)}{T_0} \boldsymbol{u}_p^T \boldsymbol{a}_k \int_{-\infty}^{+\infty} G(t) e^{j2\pi\frac{p}{\Delta T}t} e^{-j\omega t} dt \\ &= \sum_{p=-\infty}^{\infty} c_p G(\omega - \frac{2\pi p}{\Delta T})\end{aligned} \quad (21)$$

where $G(t)$ and $G(\omega)$ are the gate function and the its Fourier transform (FT), $c_p$ is a coefficient, given by

$$G(t) = \begin{cases} 1, & 0 < t \leq T_0 \\ 0, & \text{else}\end{cases} \quad (22)$$

$$G(\omega) = T_0 \text{Sa}(\omega T_0/2) e^{-j\omega T_0/2} \quad (23)$$

$$c_p = \boldsymbol{u}_p^T \boldsymbol{A} s(t_0)/T_0 \quad (24)$$

$$\boldsymbol{A} = [\boldsymbol{a}_1, \boldsymbol{a}_2, \cdots, \boldsymbol{a}_K] \quad (25)$$

$$\boldsymbol{s}(t_0) = [s_1(t_0), s_2(t_0), \cdots, s_K(t_0)]^T \quad (26)$$

$G(\omega - 2\pi p/\Delta T)$ is the shift of $G(\omega)$ in the frequency domain with different $p$, thus the FTFT of $x(t)$ can be expressed as the superposition of $G(\omega)$ on different frequency points with different coefficient $c_p$. The impulse interval of $G(\omega - 2\pi p/\Delta T)$ and the zero of $G(\omega)$ are $2\pi p/\Delta T$ and $2\pi k_0/T_0$ respectively, both $p$ and $k_0$ are integers. When $T_0 = k_0 \Delta T$, the maximum peak of $G(\omega - 2\pi p/\Delta T)$ with different $p$ are not affected by each other, shown in Fig. 3. (a). In this case, it can be seen from (21) that the maximum peaks of the "Sa" function contains the DOA information of the incident signals, and the frequency points correspond to these maximum peaks can be selected to carry out subsequent signal processing. In the following, $T_0 = k_0 \Delta T$ is always kept.

We note that the FTFT of $x(t)$ can be expressed in another way from (21): the output of each metasurface element is received separately, and the obtained multi-channel output is remodulated with $\boldsymbol{u}_p$ as the weight. The sum of the remodulation results is equivalent to the FTFT of $x(t)$, shown in Fig. 3. (b). The weight of the $p$-th pattern is controlled by the periodical coding pattern, so it can be used as a degree of freedom for optimization. This is the focus of the future work.

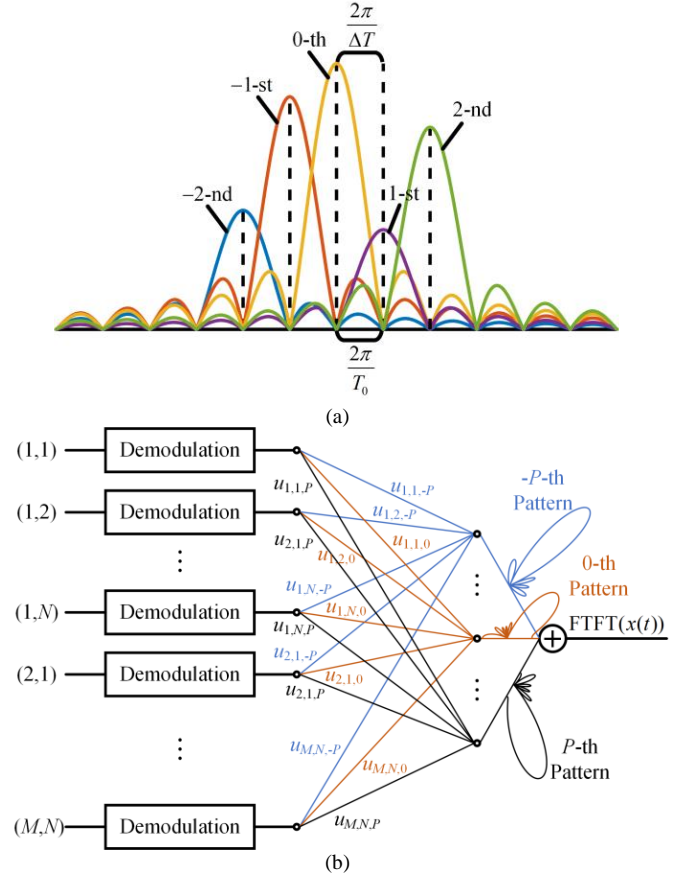

Fig. 3. The interpretation of $\text{FTFT}(x(t))$. (a) The superposition of $|G(\omega)|$ on different frequency points with different coefficient $c_p$ ($p=0, \pm 1, \pm 2$) when $T_0 = k_0\Delta T$, $k_0=1$. (b) The diagram of multi-channel output and remodulation.

Let the index of the maximum frequency point be $P$ and select frequency points corresponding to $2\pi p/\Delta T$, $p = 0, \pm 1, \cdots, \pm P$, a total of $2P+1$ frequency points can be selected. According to (21), the $p$-th frequency point can be written as

$$y_p(t_0) = \boldsymbol{u}_p^T \boldsymbol{A} \boldsymbol{s}(t_0) + n_p(t_0) \quad (27)$$

where $n_p(t_0)$ is the noise at the corresponding frequency point, then the frequency single-snapshot $\boldsymbol{y}(t_0)$ can be written as

$$\boldsymbol{y}(t_0) = [y_{-P}(t_0), \cdots, y_0(t_0), \cdots, y_P(t_0)]^T = \boldsymbol{U}\boldsymbol{A}\boldsymbol{s}(t_0) + \boldsymbol{n}(t_0) \quad (28)$$

where

$$\boldsymbol{U} = [\boldsymbol{u}_{-P}, \cdots, \boldsymbol{u}_0, \cdots, \boldsymbol{u}_P]^T \quad (29)$$

$$\boldsymbol{n}(t_0) = [n_{-P}(t_0), \cdots, n_0(t_0), \cdots, n_P(t_0)]^T \quad (30)$$

Through multiple sampling, multiple snapshots in the frequency domain can be obtained. When the interval of two samples is greater than the correlation time of the incident signals, the multi-snapshot model in the frequency domain can be obtained, that is

$$Y = UAS + N \tag{31}$$

where

$$Y = [y(t_0), y(t_1), \cdots, y(t_{I-1})] \tag{32}$$
$$S = [s(t_0), s(t_1), \cdots, s(t_{I-1})] \tag{33}$$
$$N = [n(t_0), n(t_1), \cdots, n(t_{I-1})] \tag{34}$$

in which $I$ is the number of frequency snapshots. So far, we have established single-snapshot and multi-snapshot model in frequency domain. It is worth mentioning that conventional DOA estimation algorithms for single-snapshot and multi-snapshot model can both be used rely on (28) and (31).

## III. ALGORITHM

Based on the signal model derived in section II, this part performs DOA estimation through PS technique. Firstly, appropriate sampling time is used to select frequency points. Since the received analog signal becomes discrete signal after a series of processing, fast Fourier transform (FFT) is used to replace the FTFT. Then, PS technique is applied to perform DOA estimation. This part performs 1-D estimation first, and then generalizes it to 2-D estimation.

Suppose that the noise received by each element of the metasurface is white Gaussian noise with a mean of zero and a variance of $\sigma^2$, then the noise received by the entire metasurface is white Gaussian noise with a mean of zero and a variance of $MN\sigma^2$. Assuming that the total number of sampling points is $Q$, after FFT and normalization, the $n_p(t_0)$ in (27) is still white Gaussian noise with a mean of zero and a variance of $MN\sigma^2/Q$.

The single-snapshot model shown in (28) is sufficient for PS, so the following discussions are based on this single-snapshot model. The performance of DOA estimation can be improved by using the multi-snapshot model, which we verify section IV.

### A. Frequency points selection method based on FFT

Because FFT is used in place of FTFT when processing discrete data, the phenomenon of period extension occurs in the frequency domain with sampling frequency $f_s$ as the period. We note that $T_0 = k_0 \Delta T$ can only ensure the frequency points are not affected by the "Sa" functions in the same period. Since FTFT($x(t)$) is infinite in the frequency domain, no matter what $f_s$ is chosen, the folding caused by the period extension phenomenon is inevitable after replacing FTFT with FFT, shown in Fig. 4 (The frequency domain has not been discretized). In order to reduce the impact of the extended period, $f_s \gg P/\Delta T$ and $f_s = z/\Delta T$ where $z$ is an integer should be satisfied. The spectrum shift of the results after FFT is carried out to simplify the process of frequency points selection.

$T_0 f_s$ points can be sampled in the sampling time $T_0$, so the resolution in the frequency domain is $2\pi/T_0$. The received signal is a baseband signal after demodulation; therefore, the center frequency is zero and the corresponding index is $T_0 f_s / 2 + 1$. In order to obtain (28), the frequency points which correspond to $2\pi p/\Delta T$ should be selected. Since $(2\pi p/\Delta T)/(2\pi/T_0) = k_0 p$, the indexes of the frequency points which need to be selected are

$$\text{index}_p = T_0 f_s / 2 + 1 + k_0 p \quad p = 0, \pm 1, \cdots, \pm P \tag{35}$$

### B. PS technique

#### 1) 1-D estimation

Assume $\phi_k$ is known, and it is denoted as $\phi_0$. It can be seen that when the number of selected frequency points is greater than $MN$, the multi-channel information of different elements in the metasurface can be recovered through (28) as

$$\begin{aligned} y(t_0) &= (U^H U)^{-1} U^H (UAs(t_0) + n(t_0)) \\ &= As(t_0) + (U^H U)^{-1} U^H n(t_0) \end{aligned} \tag{36}$$

For a metasurface whose structure is determined, $\tau_{m,n}$ can be computed, so define the compensation matrix $J_1$ as

$$J_1 = \text{diag}\{e^{-j\omega_0 \tau_{1,1}}, \cdots, e^{-j\omega_0 \tau_{1,N}}, \cdots, e^{-j\omega_0 \tau_{M,1}}, \cdots, e^{-j\omega_0 \tau_{M,N}}\} \tag{37}$$

The $M \times MN$ dimensional PS matrix $J_{2,l}$ is a block diagonal matrix consisting of $M$ $l$-th PS weights $j_l$, defined as

$$j_l = [j_{l,1}, j_{l,2}, \cdots, j_{l,N}] \tag{38}$$
$$J_{2,l} = \text{blkdiag}\{\underbrace{j_l, j_l, \cdots, j_l}_{M}\} \tag{39}$$

Define the $g_{l,k}$ and $b_k$ as

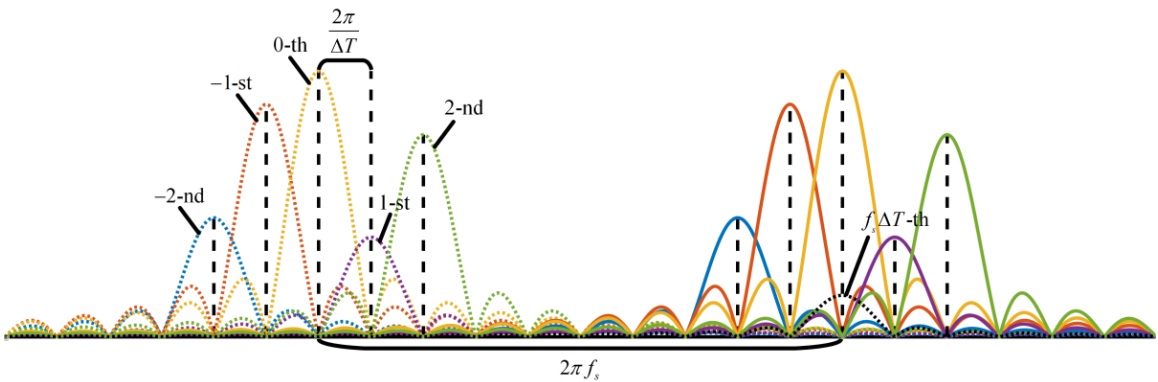

Fig. 4. Schematic diagram of period extension, only one extended period has been drawn. ($p=0, \pm1, \pm2$), the black dashed line represents the effect of the "Sa" function in the extended period on the original period at $p=0$.

$$g_{l,k} = \sum_{n=1}^{N} j_{l,n} e^{j\omega_0 (n-(N+1)/2) d \sin\phi_0 \cos\theta_k / c} \quad (40)$$

$$\boldsymbol{b}_k = \begin{bmatrix} e^{j\omega_0 (1-(M+1)/2) d \sin\phi_0 \sin\theta_k / c} \\ e^{j\omega_0 (2-(M+1)/2) d \sin\phi_0 \sin\theta_k / c} \\ \vdots \\ e^{j\omega_0 (M-(M+1)/2) d \sin\phi_0 \sin\theta_k / c} \end{bmatrix} \quad (41)$$

then perform PS technique, we can get

$$\begin{aligned} \boldsymbol{y}_l(t_0) &= \boldsymbol{J}_{2,l} \boldsymbol{J}_1 (\boldsymbol{A}\boldsymbol{s}(t_0) + (\boldsymbol{U}^H \boldsymbol{U})^{-1} \boldsymbol{U}^H \boldsymbol{n}(t_0)) \\ &= \boldsymbol{B} \boldsymbol{\Gamma}_l \boldsymbol{s}(t_0) + \boldsymbol{J}_{2,l} \boldsymbol{J}_1 (\boldsymbol{U}^H \boldsymbol{U})^{-1} \boldsymbol{U}^H \boldsymbol{n}(t_0) \end{aligned} \quad (42)$$

where

$$\boldsymbol{B} = [\boldsymbol{b}_1, \boldsymbol{b}_2, \cdots, \boldsymbol{b}_K] \quad (43)$$

$$\boldsymbol{\Gamma}_l = \mathrm{diag}\{g_{l,1}, g_{l,2}, \cdots, g_{l,K}\} \quad (44)$$

It can be seen that $g_{l,k}$ is equivalent to beamforming by the $N$ elements in row $m$, the schematic diagram of two different patterns synthesized by different PS matrix $\boldsymbol{J}_{2,l}$ is shown in Fig. 5.

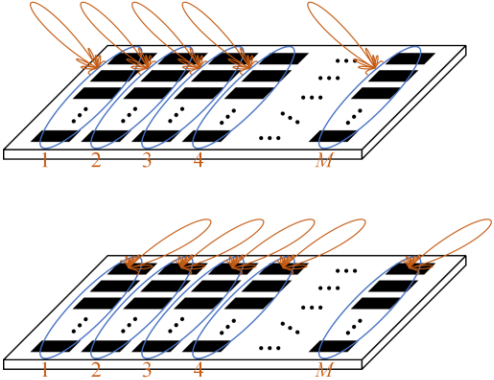

Fig. 5. Schematic diagram of two different patterns synthesized by different PS matrix in 1-D estimation.

By choosing different weights to synthesize different patterns, phase perturbation can be realized to process coherent signals. Suppose a total of $L$ weights are selected, and PS technique is performed respectively, then the frequency covariance matrix is obtained through

$$\boldsymbol{R}_{yy} = \frac{1}{L} \sum_{l=1}^{L} \boldsymbol{y}_l(t_0) \boldsymbol{y}_l^H(t_0) \quad (45)$$

(45) can be rewritten as

$$\boldsymbol{R}_{yy} = \boldsymbol{B} \boldsymbol{R}_{yy}' \boldsymbol{B}^H + \boldsymbol{R}_{nn} \quad (46)$$

$$\boldsymbol{R}_{yy}' = \frac{1}{L} \sum_{l=1}^{L} \boldsymbol{\Gamma}_l \boldsymbol{s}(t_0) \boldsymbol{s}^H(t_0) \boldsymbol{\Gamma}_l^H \quad (47)$$

$$\boldsymbol{R}_{nn} = \frac{MN}{LQ} \sigma^2 \sum_{l=1}^{L} \boldsymbol{J}_{2,l} \boldsymbol{J}_1 (\boldsymbol{U}^H \boldsymbol{U})^{-1} \boldsymbol{J}_1^H \boldsymbol{J}_{2,l}^H \quad (48)$$

Let

$$\boldsymbol{D}_{\mathrm{sum}} = \sum_{l=1}^{L} \boldsymbol{J}_l^2 \boldsymbol{J}^1 (\boldsymbol{U}^H \boldsymbol{U})^{-1} \boldsymbol{J}_1^H \boldsymbol{J}_{2,l}^H \quad (49)$$

$$\boldsymbol{g}_l = [g_{l,1}, g_{l,2}, \cdots, g_{l,K}]^T \quad (50)$$

$$\boldsymbol{H} = \mathrm{diag}\{s_1(t_0), s_2(t_0), \cdots, s_K(t_0)\} \quad (51)$$

$$\boldsymbol{\Delta} = [\boldsymbol{g}_1, \boldsymbol{g}_2, \cdots, \boldsymbol{g}_L] \quad (52)$$

First, $\boldsymbol{R}_{yy}$ is whitened as

$$\boldsymbol{R}_{yy} = \boldsymbol{D}_{\mathrm{sum}}^{-0.5} \boldsymbol{R}_{yy} (\boldsymbol{D}_{\mathrm{sum}}^{-0.5})^H \quad (53)$$

and $\boldsymbol{R}_{yy}$ can be rewritten as

$$\boldsymbol{R}_{yy} = \frac{1}{L} \boldsymbol{H} (\sum_{l=1}^{L} \boldsymbol{\delta}_l \boldsymbol{\delta}_l^H) \boldsymbol{H}^H = \frac{1}{L} (\boldsymbol{H}\boldsymbol{\Delta})(\boldsymbol{H}\boldsymbol{\Delta})^H \quad (54)$$

It can be concluded that $\mathrm{rank}(\boldsymbol{R}_{yy}) = \mathrm{rank}(\boldsymbol{H}\boldsymbol{\Delta})$ when $L \geq K$, in other words, the rank of the frequency covariance matrix can be recovered by the PS technique. For the multi-snapshot model shown in (15), this conclusion is also applicable, this is because $\mathrm{rank}(\boldsymbol{S}\boldsymbol{S}^H) = 1$ and $\boldsymbol{S}\boldsymbol{S}^H$ can be decomposed into $\boldsymbol{\eta}\boldsymbol{\eta}^H$ where $\boldsymbol{\eta}$ is a $K \times 1$ dimensional vector with non-zero elements, then let $\boldsymbol{H} = \mathrm{diag}\{\eta_1, \eta_2, \cdots, \eta_K\}$, the same conclusion can be reached.

Next, MUSIC is performed to get the DOA information of the incident signals through $\boldsymbol{R}_{yy}$, the details are given below.

The eigenvalue decomposition of $\boldsymbol{R}_{yy}$ yields $M$ eigenvalues, and the eigenvalues are arranged from large to small, i.e., $\lambda_1 \geq \lambda_2 \geq \cdots \geq \lambda_M$. The eigenvectors $\boldsymbol{v}_1, \boldsymbol{v}_2, \cdots, \boldsymbol{v}_K$ which correspond to the first $K$ eigenvalues span the signal subspace. The eigenvectors $\boldsymbol{v}_{K+1}, \boldsymbol{v}_{K+2}, \cdots, \boldsymbol{v}_M$ which correspond to the rest of the eigenvalues span the noise subspace $\boldsymbol{V}$, $\boldsymbol{V} = [\boldsymbol{v}_{K+1}, \boldsymbol{v}_{K+2}, \cdots, \boldsymbol{v}_M]$. Select the range of interest angles, and search peaks according to (55), the DOAs corresponding to the spectral peaks are the estimation results $\theta_{est}$.

$$P = |\boldsymbol{b}_s^H \boldsymbol{D}_{\mathrm{sum}}^H \boldsymbol{V} \boldsymbol{V}^H \boldsymbol{D}_{\mathrm{sum}} \boldsymbol{b}_s|^{-1} \quad (55)$$

$$\boldsymbol{b}_s = \begin{bmatrix} e^{j\omega_0 (1-(M+1)/2) d \sin\phi_0 \sin\theta_s / c} \\ e^{j\omega_0 (2-(M+1)/2) d \sin\phi_0 \sin\theta_s / c} \\ \vdots \\ e^{j\omega_0 (M-(M+1)/2) d \sin\phi_0 \sin\theta_s / c} \end{bmatrix} \quad (56)$$

where $\theta_s$ represents the azimuth which are selected during spectral peak search.

*2) 2-D estimation*

When 2-D estimation is needed, the $N_{\mathrm{sub}}$ elements in row $m$ are used to beamforming, and the PS matrix $\boldsymbol{J}_{2,l}$ needs to be changed to the following form

$$\bar{\boldsymbol{j}}_l = \begin{bmatrix} \bar{j}_{l,1}, \bar{j}_{l,2}, \cdots, \bar{j}_{l,N_{\mathrm{sub}}} \\ \quad \bar{j}_{l,1}, \bar{j}_{l,2}, \cdots, \bar{j}_{l,N_{\mathrm{sub}}} \\ \quad \quad \vdots \\ \quad \quad \bar{j}_{l,1}, \bar{j}_{l,2}, \cdots, \bar{j}_{l,N_{\mathrm{sub}}} \end{bmatrix} \quad (57)$$

$$\bar{\boldsymbol{J}}_{l,2} = \mathrm{blkdiag}\{\underbrace{\bar{\boldsymbol{j}}_l, \bar{\boldsymbol{j}}_l, \cdots, \bar{\boldsymbol{j}}_l}_{M}\} \quad (58)$$

where $\bar{\boldsymbol{j}}_l$ is $(N-N_{sub}+1)\times N$ dimension, $\bar{\boldsymbol{J}}_{l,2}$ is $M(N-N_{sub}+1)\times MN$ dimension.

Then the whitening matrix $\bar{\boldsymbol{D}}_{sum}$, the covariance matrix $\bar{\boldsymbol{R}}_{yy}$ and the corresponding noise subspace $\bar{\boldsymbol{V}}$ is obtained by the same steps as (49), (42), (45) and (53), and the spectral peaks are searched by

$$\bar{P}=|(\boldsymbol{b}_s\otimes\boldsymbol{c}_s)^H\bar{\boldsymbol{D}}_{sum}^H\bar{\boldsymbol{V}}\bar{\boldsymbol{V}}^H\bar{\boldsymbol{D}}_{sum}(\boldsymbol{b}_s\otimes\boldsymbol{c}_s)|^{-1} \quad (59)$$

$$\boldsymbol{c}_s=\begin{bmatrix}1\\ e^{j\omega_0 d\sin\phi_s\cos\theta_s/c}\\ \vdots\\ e^{j\omega_0(N-N_{sub})d\sin\phi_s\cos\theta_s/c}\end{bmatrix} \quad (60)$$

The schematic diagram of different patterns synthesized by different PS matrix $\bar{\boldsymbol{J}}_{l,2}$ in 2-D estimation is shown in Fig. 6.

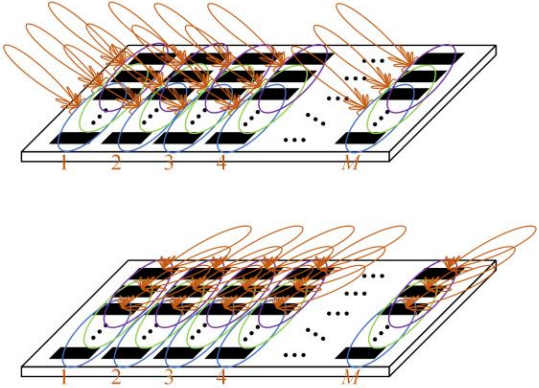

Fig. 6. Schematic diagram of two different patterns synthesized by different PS matrix in 2-D estimation.

## IV. SIMULATION RESULTS

In this part, the frequency spectrum of the received signal after periodical coding and the spatial spectrum are presented with an example, then the influence of different factors on PR and RMSE of the proposed metasurface pattern smoothing (MS-PS) method is analyzed. Besides, this part also compares the proposed method with the existing methods [21]-[24].

### A. Frequency spectrum and spatial spectrum of the received signal after periodical coding

The frequency spectrum and spatial spectrum of the received signal after periodical coding are simulated. The main simulation parameters are listed in Table I. After quadrature demodulation, the spectral peaks generated by periodical coding are all near zero frequency. The receiving antenna is set at $(0,0,-2d)$. 800 points are sampled in a periodical coding period $\Delta T$, which indicates that a frequency snapshot is constructed by every $800k_0$ sampling points.

The simulation results of frequency spectrum, 1-D spatial spectrum and 2-D spatial spectrum are shown in Fig. 7. (a), Fig. 7. (b) and Fig. 7. (c) respectively. According to the derivation and analysis above, different spectral peaks contain the spatial information of the incident signals which can be used for DOA estimation. The spatial spectrums in Fig. 7. (b) and Fig. 7. (c) prove the effectiveness of the proposed method.

The performance of the proposed method is related to the number of frequency snapshots $I$, the number of periodical coding periods $k_0$ in each snapshot, the index of the maximum frequency point $P$, the number of PS weights $L$ and $SNR$. Therefore, next part presents performance curves in different cases and compares with the existing methods.

TABLE I
PARAMETERS FOR SECTION IV. A

| Symbol | Quantity | Values |
|---|---|---|
| $N$ | The number of columns | 6 |
| $M$ | The number of rows | 5 |
| $f_0$ | Carrier frequency | 1 GHz |
| $f_s$ | Sampling frequency | 50 MHz |
| $\Delta T$ | Periodical coding period | $1.6\times 10^{-5}$ s |
| $I$ | The number of frequency snapshots | 5 |
| $L$ | The number of PS weights | 5 |
| $k_0$ | The number of $\Delta T$ in each snapshot | 2 |
| $P$ | The index of the maximum frequency point | 15 |
| $\theta_0$ | Incident angles | -22°, 12° & (-36°, 20°), (42°, 45°) |
| $SNR$ | Signal to noise ratio | 0 dB |

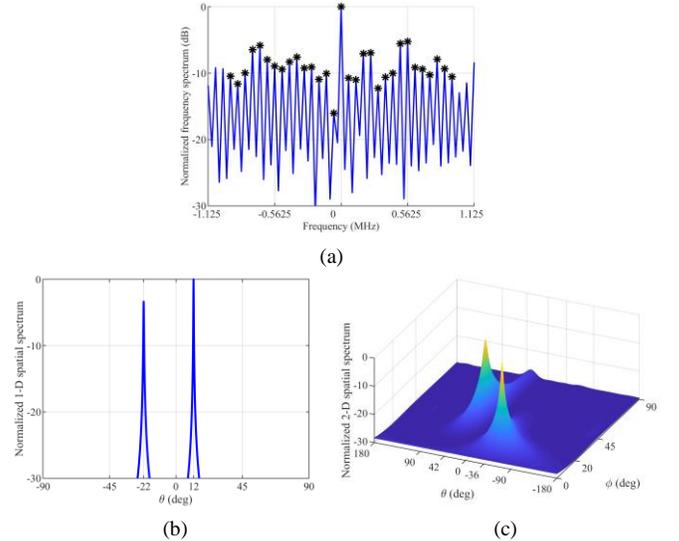

Fig. 7. Spectrums. (a) Frequency spectrum, the black markers represent the selected frequency points. (b) 1-D spatial spectrum. (c) 2-D spatial spectrum ($N_{sub}=4$).

### B. Performance analysis

The influence of different factors on PR and RMSE of the proposed method are analyzed in this part. Besides, this part also compares the proposed method with the single sensor compressive sensing (SSCS) method in [21], the nonuniformly periodic time modulation (NPTM) method in [22] and the space-time-coding digital metasurface direction finding (STCM-DF) method in [23] in the case of single or multiple incident signals. The performance curves shown are the averaged results of 500 independent trials, and the PR and RMSE are calculated by

$$\text{PR}=\frac{1}{500}\sum_{n=1}^{500}\delta_{Q_n-K}(\delta_{\left\lceil\sum_{k=1}^{K}|\theta_{est}^{k,n}-\theta_k|/(2K)-1\right\rceil}+\delta_{\sum_{k=1}^{K}|\theta_{est}^{k,n}-\theta_k|/(2K)-1}) \quad (61)$$

$$\text{RMSE}=\sqrt{\frac{1}{500K}\sum_{n=1}^{500}\sum_{k=1}^{K}(\theta_{est}^{k,n}-\theta_k)^2} \quad (62)$$

where $\delta_x$ is the Dirac delta function (that is, $\delta_x = 1$ if $x = 0$, and $\delta_x = 0$ if $x \neq 0$), $Q_n$ is the number of the main peaks found in the $n$-th trial, $\theta_{est}^{k,n}$ represents the DOA of the $k$-th incident signal estimated from the $n$-th independent trial, $\lceil x \rceil$ denotes the smallest integer greater than $x$. It should be pointed out that when the PR is low, only a few times in 500 independent trials can distinguish two signals, so the calculated RMSE loses its statistical properties, it is necessary to increase the number of trials to make RMSE more mathematically significant in this case. In order to simplify the calculation, only 1-D estimation is carried out, i.e., $\phi_k = 90°$. When PR=0, RMSE is written as $180°$. The derivation of Cramér-Rao Bounds (CRB) for the corresponding signal model (31) is detailed in the appendix.

*1) PR and RMSE of multiple incident signals*

This part analyzes the PR and the RMSE under multiple incident signals. Since the STCM-DF method can only estimate a single incident signal, the comparison with this method is given in the next part which takes a single incident signal as an example.

In the first example, the number of frequency snapshots $I$ on DOA estimation is studied. Most of the simulation parameters are the same as IV. A and the different parameters are listed in Table II. In the SSCS method, 8 elements are used, and the number of rows in the code matrix is 20. In the NPTM method, the type of metasurface used is consistent with the proposed method, and the duty ratio is 0.2, then MUSIC algorithm is used for DOA estimation. According to IV.A, each frequency snapshot requires $800k_0$ sampling points, therefore, in order to keep the time domain sampling points consistent, $800k_0 I$ sampling points are used for DOA estimation in the SSCS method. The two incident signals are incoherent. The SSCS method and the NPTM method are simulated in the same way unless otherwise stated in the following simulations. The PR and RMSE curves are shown in Fig. 8. (a) and Fig. 8. (b) respectively.

TABLE II
DIFFERENT PARAMETERS FOR SECTION IV. B *1)*

| Symbol | Quantity | Values |
|---|---|---|
| $N$ | The number of columns | 5 |
| $M$ | The number of rows | 8 |
| $I$ | The number of frequency snapshots | 1, 2, 3, 5, 10, 15, 20, 30 |
| $P$ | The index of the maximum frequency point | 20 |

Fig. 8 demonstrates that the MS-PS method has the highest PR compared with the other two methods when using a single frequency snapshot. This is because the autocorrelation matrix of the SSCS method requires more sampling points to be approximated as a diagonal matrix during calculation, while the rank of covariance matrix constructed by the NPTM method is one, and multiple incident signals cannot be estimated by MUSIC. Besides, it can be seen that the PR and RMSE of the proposed method are better than the other methods when $I$ increases, the accuracy of the estimation results can be improved by selecting a lager $I$.

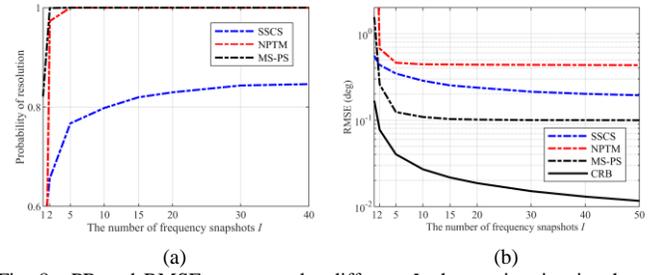

Fig. 8. PR and RMSE curves under different $I$ when estimating incoherent signals. (a) PR. (b) RMSE.

In the second example, the influence of the number of periodical coding periods $k_0$ in each snapshot on DOA estimation is studied. Most of the simulation parameters are the same as the first example except for $k_0 = 1$, 2, 3, 5, 10, 15, 20, 30 and $SNR = -10$ dB. The PR and RMSE curves are shown in Fig. 9. (a) and Fig. 9. (b) respectively.

It can be seen from Fig. 9 that with the increase of $k_0$, the performance of all methods can be improved. For the SSCS method, the increase of $k_0$ represents the increase of the number of sampling points, which means the autocorrelation matrix is closer to the diagonal matrix. For the NPTM method and the proposed MS-PS method, the increase of $k_0$ can make the peaks on the frequency spectrum more obvious, in other words, the received $SNR$ is improved through time accumulation. The increase of the sampling points will lead to the increase of the computational amount, so it is necessary to select $k_0$ appropriately.

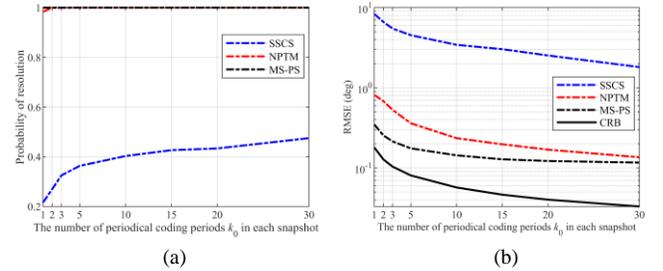

Fig. 9. PR and RMSE curves under different $k_0$ when estimating incoherent signals. (a) PR. (b) RMSE.

In the third example, the influence of $SNR$ on DOA estimation is studied. Most of the simulation parameters are the same as the first example except for $SNR$ increases from $-20$ to 20 with interval 5, coherent and incoherent cases are also compared. The PR and RMSE curves are shown in Fig. 10. (a) and Fig. 10. (b) respectively.

As can be seen from Fig. 10, the proposed method is better than the other two methods in the case of estimating two coherent or incoherent signals, however, the performance of estimating coherent signals is not as good as that of incoherent signals. In the coherent case, the other two methods failed, so their PR and RMSE curves are not drown in Fig. 10. (b). Changing the $f_s$ to $10f_s$, and calculating the RMSE of the proposed method again. The result is shown in Fig. 10. (c). With the increases of the sampling frequency, the calculation of $U$ in (36) is more accurate. It can be seen that RMSE is closer to CRB compared with Fig. 10. (b), thus verifies the correctness of the derivation in III.A.

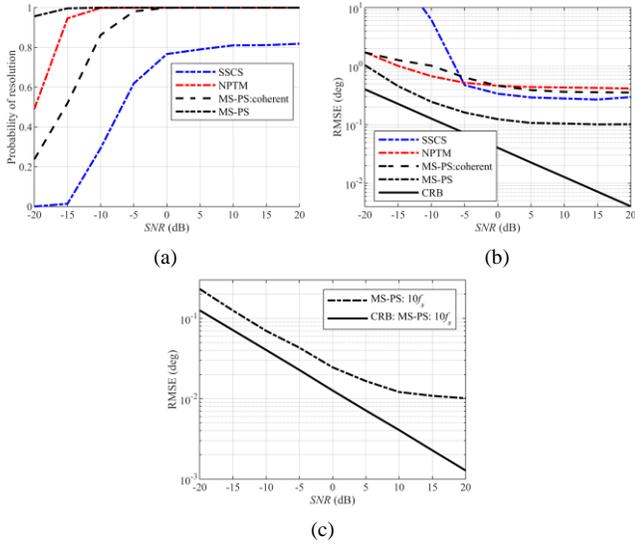

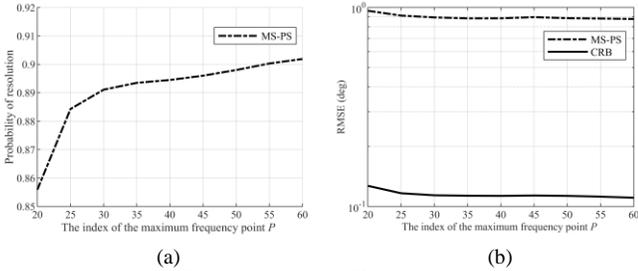

(c)

Fig. 10.  PR and RMSE curves under different *SNR* when estimating incoherent and coherent signals. (a) PR. (b) RMSE. (c)RMSE with $10f_s$.

In the fourth example, the influence of the index of the maximum frequency point $P$ on DOA estimation is studied. The other two methods are not affected by $P$, therefore, only the performance of the proposed method for estimating two coherent incident signals at different $P$ is simulated in this example. Most of the simulation parameters are the same as the first example except for $P$ increases from 20 to 60 with interval 5 and $SNR = -10$ dB. The PR and RMSE curves are shown in Fig. 11. (a) and Fig. 11. (b) respectively.

Fig. 11.  PR and RMSE curves under different $P$ when estimating coherent signals. (a) PR. (b) RMSE.

Fig. 11 shows that the performance of the MS-PS method can be slightly increased by increasing $P$. However, the complexity of the operation shown in (36) increases as $P$ increases. In practice, $P$ needs to be selected appropriately.

In the last example, the influence of the number of PS weights $L$ on DOA estimation is studied. Same as the previous example, the other two methods are not affected by $L$, only the performance of the proposed method for estimating two coherent incident signals at different $L$ is simulated. Most of the simulation parameters are the same as the first example except for $L = 2$, 5, 10, 15, 20, 40, 60 and $SNR = -10$ dB. The PR and RMSE curves are shown in Fig. 12. (a) and Fig. 12. (b) respectively.

Fig. 12 shows that the performance of the proposed MS-PS method is related to the number of PS weights $L$, and will be improved with the increase of $L$. When more than 20 weights are used, the performance increases slowly. We select the PS weights randomly in the simulations, so this is a degree of freedom that can improve the performance of the proposed method.

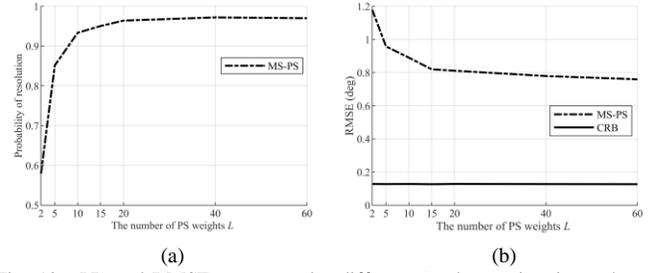

Fig. 12.  PR and RMSE curves under different $L$ when estimating coherent signals. (a) PR. (b) RMSE.

It can be concluded from the above simulations that the accuracy of DOA estimation can be improved by increasing $I$, $k_0$, $P$ and $L$. However, the computational complexity will increase with the increase of these parameters. Therefore, in practical application, appropriate parameters should be selected to satisfy different application scenarios.

*2) PR and RMSE of single incident signal*

The performance of the STCM-DF method is added to the comparison in this part in single incident scenario. In the STCM-DF method, 8 elements are still employed, and in order to keep the number of sampling points fair, $800k_0$ sampling points are used to calculate the result once, and a total of $I$ results are averaged to get the final result. The simulation experiments are the same as the simulations designed in *1)* except for the incident angle is 22°. The PR and RMSE curves under different $I$, $k_0$, $SNR$ and $P$ are shown in Fig. 13 respectively.

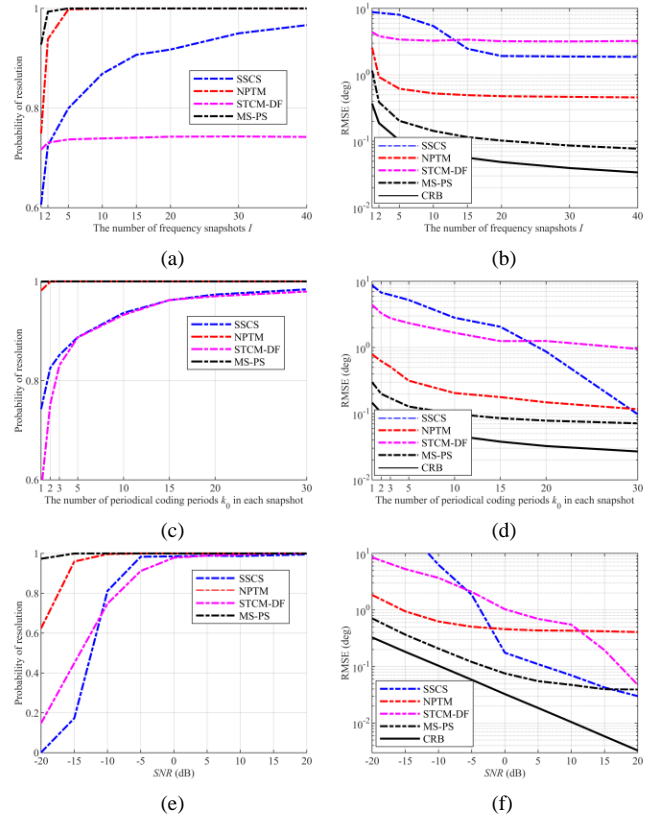

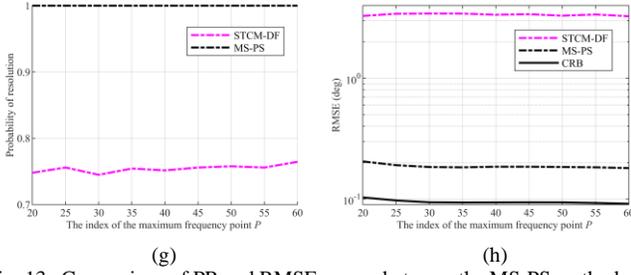

Fig. 13. Comparison of PR and RMSE curves between the MS-PS method and the existing methods under different $I$, $k_0$, $SNR$, and $P$. (a) PR under different $I$. (b) RMSE under different $I$. (c) PR under different $k_0$. (d) RMSE under different $k_0$. (e) PR under different $SNR$. (f) RMSE under different $SNR$. (g) PR under different $P$. (h) RMSE under different $P$.

It can be seen from Fig. 13 that the performance of the proposed MS-PS method is better than the STCM-DF method in all simulations. Unlike estimating two coherent signals, changes in $P$ have little effect on RMSE when estimating one signal. Besides, the decreasing trend of RMSE curves of the MS-PS method under different simulations is the same as that obtained from *1)*, and the performance of the SSCS method exceeds the proposed method only when $SNR = 20$ dB. The SSCS method is a sparse reconstruction method, its calculation time is much higher than the proposed spectrum estimation method; thus, it can be said that the proposed MS-PS method is superior to the existing methods in terms of performance and calculation time.

## V. METHOD TO IMPROVE THE PERFORMANCE OF ESTIMATION

According to (21), when there are only a finite number of gate functions in the frequency domain and $f_s > (2P+1)/\Delta T$, the folding effect will not exist, therefore; down-conversion and band-pass filtering processes can be added to the flow in Fig. 1, shown in Fig. 14.

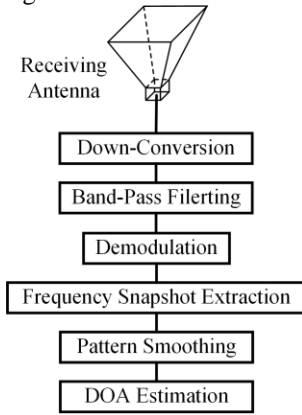

Fig. 14. The flow chart for the method to improve the performance of estimation.

When $f_s \gg (2P+1)/\Delta T$, the passband of the bandpass filter can be larger, making the frequency response smoother over the desired frequency range. We assume that in an ideal case, that is, all gate functions outside the passband are filtered out. The normalized frequency spectrum and RMSE are shown in Fig. 15. (a) and Fig. 15. (b) respectively under the assumption of the third simulation example in IV. B. *1)*. It can see that this method removes the effect of folding under the ideal condition. Because the number of equivalent array elements decreases after PS, there is still a gap between the RMSE and CRB. The effect of the band-pass filter on the improved method needs to be further studied in practical applications.

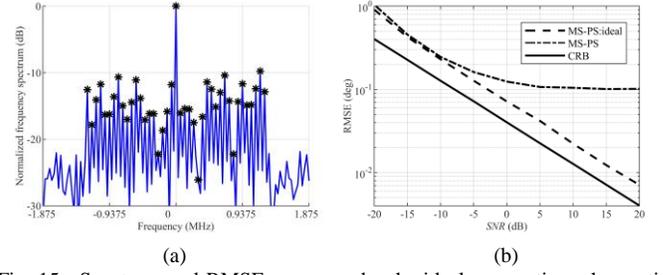

Fig. 15. Spectrum and RMSE curves under the ideal assumption when estimating coherent signals. (a) Frequency spectrum, the black markers represent the selected frequency points. (b) RMSE.

## VI. CONCLUSION

In this paper, a DOA estimation method based on periodical coding metasurface is proposed, which needs to select appropriate frequency points in the frequency domain to reconstruct frequency snapshots, besides, the use of PS technique enables the proposed method to process coherent signals, and the use of metasurface can achieve single channel reception, which avoids the appearance of channel mismatch errors in traditional array system. The effectiveness of the proposed method is verified by mathematical simulation, and the statistical characteristics are compared with the existing methods. The simulation results show that the performance of the proposed method is better than the existing methods, and the accuracy of the proposed method can be improved by selecting appropriate parameters. The design of periodical coding pattern and the selection of PS weights are the degrees of freedom which can improve the performance, and are worth of further study.

## APPENDIX

Assume that $\{s(t_z)\}$ $(z = 0, 1, \cdots, I-1)$ is a deterministic, unknown sequence, this appendix deduces the CRB based on the multi-snapshot model in (31). Under this assumption, let $\mathbf{y} = \text{vec}(\mathbf{Y})$, then the $(2P+1)I \times 1$ dimensional mean matrix $\boldsymbol{\mu}$ and the $(2P+1)I \times (2P+1)I$ dimensional covariance matrix $\mathbf{Z}$ of $\mathbf{y}$ are given by

$$\boldsymbol{\mu} = [(\mathbf{UAs}(t_0))^\mathrm{T}, (\mathbf{UAs}(t_1))^\mathrm{T}, \cdots, (\mathbf{UAs}(t_{I-1}))^\mathrm{T}]^\mathrm{T} \quad \text{(A.1)}$$

$$\mathbf{Z} = MN\sigma^2 \mathbf{I}_{(2P+1)I} / Q \quad \text{(A.2)}$$

where $\mathbf{I}_k$ represents the $k \times k$ dimensional unit matrix.

The parameter vector to be estimated is given by

$$\boldsymbol{\Psi} = [\underbrace{\boldsymbol{\theta}^\mathrm{T}, \boldsymbol{\phi}^\mathrm{T}}_{\boldsymbol{\varepsilon}^\mathrm{T}}, \text{Re}(\text{vec}(\mathbf{S}))^\mathrm{T}, \text{Im}(\text{vec}(\mathbf{S}))^\mathrm{T}, \sigma^2]^\mathrm{T} \quad \text{(A.3)}$$

where $\boldsymbol{\theta}^\mathrm{T} = [\theta_1, \theta_2, \cdots, \theta_K]$, $\boldsymbol{\phi}^\mathrm{T}$ is defined the same as $\boldsymbol{\theta}^\mathrm{T}$, Re($\cdot$) and Im($\cdot$) represent taking the real part and the imaginary part respectively. The $(l_1, l_2)$-th element of the Fisher information matrix is

$$F_{l_1, l_2} = \text{tr}\left\{\mathbf{Z}^{-1} \frac{\partial \mathbf{Z}}{\partial \Psi_{l_1}} \mathbf{Z}^{-1} \frac{\partial \mathbf{Z}}{\partial \Psi_{l_2}}\right\} + 2\text{Re}\left\{\frac{\partial \boldsymbol{\mu}^\mathrm{H}}{\partial \Psi_{l_1}} \mathbf{Z}^{-1} \frac{\partial \boldsymbol{\mu}}{\partial \Psi_{l_2}}\right\} \quad \text{(A.4)}$$

By calculation, we can get

$$\frac{\partial \boldsymbol{\mu}}{\partial \boldsymbol{\varepsilon}^{\mathrm{T}}} = \underbrace{(\mathbf{1}_{I\times 1} \otimes \boldsymbol{U}\boldsymbol{\Theta}) \odot (\mathbf{1}_{1\times 2} \otimes (\boldsymbol{S}^{\mathrm{T}} \otimes \mathbf{1}_{(2P+1)\times 1}))}_{\boldsymbol{\Lambda}_1} \quad (\mathrm{A.5})$$

$$\frac{\partial \boldsymbol{\mu}}{\partial \mathrm{Re}(\mathrm{vec}(\boldsymbol{S}))^{\mathrm{T}}} = \underbrace{\boldsymbol{I}_I \otimes \boldsymbol{U}\boldsymbol{A}}_{\boldsymbol{\Lambda}_2} \quad (\mathrm{A.6})$$

$$\frac{\partial \boldsymbol{\mu}}{\partial \mathrm{Im}(\mathrm{vec}(\boldsymbol{S}))^{\mathrm{T}}} = \mathrm{j}\boldsymbol{\Lambda}_2 \quad (\mathrm{A.7})$$

$$\frac{\partial \boldsymbol{\mu}}{\partial \sigma^2} = \mathbf{0}_{(2P+1)I\times 1} \quad (\mathrm{A.8})$$

$$\boldsymbol{\Theta} = \left[ \frac{\partial \boldsymbol{A}(:,1)}{\partial \theta_1}, \cdots, \frac{\partial \boldsymbol{A}(:,K)}{\partial \theta_K}, \frac{\partial \boldsymbol{A}(:,1)}{\partial \phi_1}, \cdots, \cdots, \frac{\partial \boldsymbol{A}(:,K)}{\partial \phi_K} \right] \quad (\mathrm{A.9})$$

where $\mathbf{1}_{k_1 \times k_2}$ ($\mathbf{0}_{k_1 \times k_2}$) represents the $k_1 \times k_2$ dimensional vector whose elements are all 1 (0), $\otimes$ and $\odot$ represent Kronecker product and Hadamard product respectively, $\boldsymbol{A}(:,k)$ represents the $k$-th column of $\boldsymbol{A}$. The dimensions of $\boldsymbol{\Lambda}_1$ and $\boldsymbol{\Lambda}_2$ are $(2P+1)I \times 2K$ and $(2P+1)I \times KI$. Then the complete Fisher information matrix can be written as

$$\boldsymbol{F} = \begin{bmatrix} \dfrac{2Q}{MN\sigma^2}\boldsymbol{\Lambda}_3 & \mathbf{0}_{(2K+2KI)\times 1} \\ \mathbf{0}_{1\times(2K+2KI)} & \dfrac{(2P+1)I}{\sigma^4} \end{bmatrix} \quad (\mathrm{A.10})$$

$$\boldsymbol{\Lambda}_3 = \mathrm{Re}\left\{[\boldsymbol{\Lambda}_1, \boldsymbol{\Lambda}_2, \mathrm{j}\boldsymbol{\Lambda}_2]^{\mathrm{H}} [\boldsymbol{\Lambda}_1, \boldsymbol{\Lambda}_2, \mathrm{j}\boldsymbol{\Lambda}_2]\right\} \quad (\mathrm{A.11})$$

and the inverse of the $2K \times 2K$ dimensional matrix in the upper left corner is the CRB matrix of the desired parameters which need to be estimated.

Let

$$\boldsymbol{\Lambda}_4 = \begin{bmatrix} \boldsymbol{I}_{2K} & & \\ -\mathrm{Re}\{\boldsymbol{\Lambda}_2^{\dagger}\boldsymbol{\Lambda}_1\} & \boldsymbol{I}_{KI} & \\ -\mathrm{Im}\{\boldsymbol{\Lambda}_2^{\dagger}\boldsymbol{\Lambda}_1\} & & \boldsymbol{I}_{KI} \end{bmatrix} \quad (\mathrm{A.12})$$

$$\boldsymbol{\Lambda}_2^{\dagger} = (\boldsymbol{\Lambda}_2^{\mathrm{H}}\boldsymbol{\Lambda}_2)^{-1}\boldsymbol{\Lambda}_2^{\mathrm{H}} \quad (\mathrm{A.13})$$

$$\boldsymbol{\Pi}_{\boldsymbol{\Lambda}_2}^{\perp} = \boldsymbol{I}_{(2P+1)I} - \boldsymbol{\Lambda}_2 \boldsymbol{\Lambda}_2^{\dagger} \quad (\mathrm{A.14})$$

then we diagonalize $\boldsymbol{\Lambda}_3$ as

$$\boldsymbol{\Lambda}_4^{\mathrm{T}} \boldsymbol{\Lambda}_3 \boldsymbol{\Lambda}_4 = \mathrm{Re}\left\{ \begin{bmatrix} \boldsymbol{\Lambda}_1^{\mathrm{H}}\boldsymbol{\Pi}_{\boldsymbol{\Lambda}_2}^{\perp}\boldsymbol{\Lambda}_1 & \mathbf{0}_{2K\times KI} & \mathbf{0}_{2K\times KI} \\ \mathbf{0}_{KI\times 2K} & \boldsymbol{\Lambda}_2^{\mathrm{H}}\boldsymbol{\Lambda}_2 & \mathrm{i}\boldsymbol{\Lambda}_2^{\mathrm{H}}\boldsymbol{\Lambda}_2 \\ \mathbf{0}_{KI\times 2K} & -\mathrm{i}\boldsymbol{\Lambda}_2^{\mathrm{H}}\boldsymbol{\Lambda}_2 & \boldsymbol{\Lambda}_2^{\mathrm{H}}\boldsymbol{\Lambda}_2 \end{bmatrix} \right\} \quad (\mathrm{A.15})$$

We only care about the $2K \times 2K$ dimensional block matrix in the upper left corner, so after the elementary row transformation and taking the $2K \times 2K$ dimensional block matrix in the upper left corner, we can get

$$\begin{aligned} \mathrm{CRB}(\boldsymbol{\varepsilon}) &= \left\{ \boldsymbol{\Lambda}_4 (\frac{2Q}{MN\sigma^2}\boldsymbol{\Lambda}_4^{\mathrm{T}}\boldsymbol{\Lambda}_3\boldsymbol{\Lambda}_4)^{-1}\boldsymbol{\Lambda}_4^{\mathrm{T}} \right\}_{2K\times 2K} \\ &= \frac{MN\sigma^2}{2Q} \mathrm{Re}\left\{ \boldsymbol{\Lambda}_1^{\mathrm{H}}\boldsymbol{\Pi}_{\boldsymbol{\Lambda}_2}^{\perp}\boldsymbol{\Lambda}_1 \right\}^{-1} \end{aligned} \quad (\mathrm{A.16})$$

Mathematically, (A.16) can also be written in the following form

$$\mathrm{CRB}(\boldsymbol{\varepsilon}) = \frac{MN\sigma^2}{2QI} \mathrm{Re}\left\{ \boldsymbol{\Omega}_1 \odot \boldsymbol{\Omega}_2 \right\}^{-1} \quad (\mathrm{A.17})$$

where $\boldsymbol{\Omega}_1 = \boldsymbol{\Theta}^{\mathrm{H}}\boldsymbol{U}^{\mathrm{H}}\boldsymbol{\Pi}_{\boldsymbol{UA}}^{\perp}\boldsymbol{U}\boldsymbol{\Theta}$, $\boldsymbol{\Omega}_2 = (\mathbf{1}_{2\times 2} \otimes \boldsymbol{R}_{SS})^{\mathrm{T}}$ and $\boldsymbol{R}_{SS} = \boldsymbol{S}\boldsymbol{S}^{\mathrm{H}} / I$.